\documentclass[10pt]{article}

\usepackage[margin=1in]{geometry}
\usepackage{graphicx}% Include figure files
\usepackage{dcolumn}% Align table columns on decimal point
\usepackage{bm}% bold math
\usepackage{amsmath}
\usepackage{amssymb}
\DeclareMathOperator{\sgn}{sgn}
\usepackage{xcolor}
\usepackage{float}
\usepackage[normalem]{ulem}
\begin{document}

\title{Emergent heavy-tailed distributions from a Markovian random walk}
\author{Henrique S. Lima\thanks{Email: hslima94@cbpf.br} \and Evaldo M. F. Curado\\[0.5em]
\small Centro Brasileiro de Pesquisas F\'{\i}sicas and National Institute of Science and Technology for Complex Systems,\\
\small Rua Dr. Xavier Sigaud 150, 22290-180, Rio de Janeiro, Brazil}
\date{}

\maketitle

\begin{abstract}
The emergence of heavy-tailed statistics in complex systems is conventionally attributed to non-local stochastic jumps or non-Markovian memory. Here, we present a one-dimensional random walk where power-law behaviors arise instead from a strictly local, discrete-time Markovian mechanism. The step length is governed by a deterministic function of the walker's position, establishing a positive feedback loop that induces strong effective correlations along the trajectories. Through analytical derivations in the continuum limit and extensive numerical simulations, we show that this rule yields a robust, non-Gaussian stationary state. The exact analytical solution is obtained in the closed form of a symmetric, Lorentz-like distribution, $\rho_{\text{st}}(x) \propto (|x|/l+r\Delta x)^{-2}$, confirming asymptotic power-law tails that decay as $|x|^{-2}$ over six decades. Furthermore, by employing the Onsager-Machlup path-integral formalism, we demonstrate that effective velocity and acceleration acquire physical meaning along the shortest fluctuation trajectories. Crucially, we find that a non-zero initial acceleration acts as the fundamental mechanism driving the walker away from the origin, ensuring both the emergence of scale-free statistics and the normalizability of the stationary distribution. This minimal pathway provides a new microscopic foundation for the widespread $-2$ power law observed across multidisciplinary complex systems. 
\end{abstract}


%\tableofcontents

\section*{Introduction}

Random walks represent a fundamental paradigm in statistical physics, providing a mathematical framework to model a vast range of phenomena, from the diffusion of particles in fluids \cite{einstein1905, chandrasekhar1943} to price fluctuations in financial markets \cite{mandelbrot1963, mantegna1995} and foraging patterns of animals \cite{viswanathan1999}. Crucially, this paradigm has also transcended physical systems, becoming a cornerstone for network science, data mining, and algorithmic optimization \cite{Xia2020}. The archetypal model, first posed in the context of a ``random walk" by Pearson \cite{pearson1905}, is the simple random walk, where a walker takes steps of fixed size in random directions. This process is inherently Markovian, meaning the next step depends only on the current position, not on the history of the trajectory \cite{spitzer_book, feller_book}. As a direct consequence of the Central Limit Theorem, the long-time probability distribution of the walker's position converges to a Gaussian distribution, with a mean-squared displacement (MSD) that grows linearly with time, $\langle x^2(t) \rangle \propto t$ \cite{bouchaud1990}.

However, a wide range of systems across physics, biology, and economics exhibit transport dynamics that deviate from this standard diffusive framework. This ``anomalous diffusion" is characterized by a power-law scaling of the MSD, $\langle x^2(t) \rangle \propto t^\alpha$, where $\alpha > 1$ corresponds to superdiffusion and $\alpha < 1$ to subdiffusion \cite{metzler2000, shlesinger1993,BeckTsallis2025}. Such behaviors typically emerge from the breakdown of the microscopic assumptions underlying the simple random walk. Traditional theoretical frameworks to model these phenomena include continuous-time random walks (CTRWs) with heavy-tailed waiting time distributions for subdiffusion \cite{montroll1965}, and Lévy flights or Lévy walks for superdiffusion \cite{shlesinger1985, Shlesingeretal1995, zaburdaev2015}. Mechanistically, alternative approaches rely on breaking standard local or stationary assumptions, either by introducing long-range spatial jumps, or by incorporating explicit path-history dependencies and aging effects that fundamentally link anomalous dynamics to generalized entropy 
formalisms~\cite{Tirnaklietal2011, HanelThurner2013}.

The emergence of heavy-tailed, power-law probability distributions is a prominent characteristic of complex systems \cite{newman2005, clauset2009, sornette2006}, suggesting scale-free organization and the absence of a well-defined characteristic scale at the asymptotic limit. Conventionally, the microscopic origin of such distributions is attributed to inherently stochastic processes, such as drawing jump lengths from a Lévy-stable distribution \cite{mandelbrot1982, metzler2000}. 

In this work, we introduce a  mechanism for generating such power-law statistics. We present a one-dimensional, discrete-time  random walk where the emergence of heavy tails is not dictated by a predefined stochastic jump distribution, but rather arises from a simple, deterministic, and position-dependent rule for the step size.
Crucially, our model is strictly Markovian, as the walker's next position is determined solely by its current displacement. Nevertheless, the deterministic dependence of the step length on the distance from the origin induces strong effective correlations in the walker's trajectory. A walker far from the origin is propelled to take even larger steps, creating a positive feedback loop that facilitates long excursions. We demonstrate, through analytical derivations and extensive numerical simulations, that this simple microscopic rule leads to a strongly non-Gaussian stationary distribution. The distribution exhibits power-law tails decaying as $\rho(x) \sim |x|^{-2}$, a functional form reminiscent of the Cauchy-Lorentz distribution, which is a stable distribution with an undefined variance \cite{feller_book}. This work connects simple, deterministic local dynamics with complex macroscopic statistical behavior, offering an alternative framework on the origins of power-law statistics in systems that may lack intrinsic long-range stochastic jumps or explicit memory \cite{hughes1995, havlin2002}. This approach may find applications in modeling systems where mobility or activity is environmentally dependent, such as diffusion in heterogeneous media \cite{benichou2014} or economic models with wealth-dependent volatility \cite{bouchaud2000}.

\section*{Model and methods}

%\subsection{The accelerated random walk}

We consider a one-dimensional  random walk where the position of the walker at a discrete time step $t$ is denoted by $n_t$. The process starts at the origin, $n_0 = 0$. The evolution of the walker's \textit{dimensionless} position is governed by the stochastic recurrence relation:
\begin{equation}
    n_{t+1} = n_t + \sigma_t s(n_t),
\end{equation}
where $\sigma_t$ is a random variable independent and identically distributed at each step, taking values $\{+1, -1\}$ with equal probability, $P(\sigma_t = \pm 1) = 1/2$. The physical position of the walker is defined via $x_t \equiv n_t \Delta x$, where $\Delta x$ represents the elementary spatial increment of the process.

The key feature of the model is the position-dependent step size, $s(n_t)$, which is defined by a piecewise function based on two positive integer parameters: a spatial threshold $l$ and a rule parameter $r$. The step size is given by:
\begin{equation}
    s(n_t) = 
    \begin{cases} 
        1 & \text{if } |n_t| \le l \\
        \lfloor |n_t|/l \rfloor + r & \text{if } |n_t| > l\,.
    \end{cases}
\end{equation}
Here, $\lfloor \cdot \rfloor$ denotes the floor function.

This rule defines two distinct dynamical regimes. For positions within the central region defined by $|n_t| \leq l$, the walker performs a standard simple random walk with a unit step size. When the walker's trajectory exceeds the threshold $l$, it enters an accelerated regime. For instance, for $l=10$ and $r=1$, the particle undergoes a standard random walk in the interval $n_t\in [-10,10]$. However, after $n_t$ surpasses this threshold ($|n_t|>10$), the particle has equal probabilities of returning to the standard rule or following the rules $n_{t+1}=n_t+2\sigma_t$, $n_{t+1}=n_t+3\sigma_t$, $\dots$, $n_{t+1}=n_t+m\sigma_t$, and so forth, where $m-1$ is the number of thresholds surpassed by the walker. In this outer region, the step size increases in discrete jumps as the walker moves farther from the origin. 

The statistical properties of the model are investigated through extensive numerical simulations. An ensemble of $N$ independent (non-interacting) walkers is simulated in parallel using a GPU in Python. Each simulation runs for a total of $N_{steps}$ time steps.

The stationary probability distribution of the walker's position, $\rho(x)$, is approximated by constructing a normalized histogram from the ensemble of final positions. Both the number of walkers $N$ and simulation time $N_{steps}$ are chosen to be sufficiently large to ensure statistical convergence and accurate characterization of the system's long-time behavior across different values of $r$ and $l$.

We simulate an ensemble of $N=2\times 10^6$ random walks over a total of $N_{steps}=5\times 10^7$ steps. We average the observables over $2.5 \times 10^7$ steps after each walk undergoes a transient of $2.5 \times 10^7$ steps. 

\section*{Results}
For simplicity, we first compute the first and second moments from the time evolution of $x_t$ for $l=10$ and $r=1$. For the first moment, we have 
\begin{equation}
    x_{t+1}-x_t=\Delta x\,\sigma_ts(x_t)=\sigma_t\left\lfloor\frac{|x_t|}{10}\right\rfloor+\Delta x\,\sigma_t\,,
\end{equation}
Thus, in the continuous limit, we have 
\begin{equation}
    \tau \frac{d}{dt}\langle x\rangle_{\sigma_t}=0,
\end{equation}
which yields a zero mean.

For the second moment in the continuous limit, we have 

\begin{equation}
    \tau \frac{d}{dt}\langle x^2 \rangle
\approx \frac{1}{100}\langle x^2\rangle+\frac{\Delta x}{5}\langle|x| \rangle+(\Delta x)^2\,. \end{equation}
For large $x$, we have $\langle x^2 \rangle\gg \langle |x| \rangle$ and $\langle x^2 \rangle\gg 1$; therefore, we conclude that the second moment also diverges exponentially as $\langle x^2 \rangle_t\sim \langle x^2 \rangle_0 e^{t/100\tau}$. In the general case, we have $\langle x\rangle=0$ and $\langle x^2 \rangle_t\sim \langle x^2 \rangle_0 e^{t/(l^2\tau)}$.

Let us now consider the generating function, for which the continuous limit corresponds to $G(k,t)=\int d\sigma\, P(\sigma) \int dx \rho(x,t) e^{ik\,x}\equiv\langle e^{ik\,x} \rangle_{\sigma,x}$, where $P(\sigma)=\frac{1}{2}(\delta(\sigma+1)+\delta(\sigma-1))$. The generating function after averaging over the values of $\sigma$ is then

\begin{equation}
    \tau \dot{G}(k,t)=\left\langle e^{ikx}\cos{ks(x)} \right\rangle_x-G(k,t)\,.
\end{equation}
By assuming $ks(x)\ll1$, we have the following approximation:
\begin{equation}
    \tau \dot{G}(k,t)\approx -\frac12k^2\left\langle s(x)^2e^{ikx} \right\rangle_x\,.
\end{equation}
Therefore,
\begin{equation}
    \tau\frac{\partial \rho}{\partial t}\approx \frac12\frac{\partial^2 }{\partial x^2}[s(x)^2\rho]\,,
\end{equation}
where, in the continuous limit $s(x)\sim \frac{|x|}{l}+r\Delta x$. This random walk admits a stationary distribution, $\rho_{st}$, for which $\frac{\partial \rho_{st}}{\partial t}=0$; therefore, we finally obtain the solution of the diffusion equation with zero probability flux, given by $\rho_{st}\propto \frac{1}{(|x|/l+r\Delta x)^2}$, which decays as $\rho_{st}\sim \rho_0 |x|^{-2}$, where $\rho_0=r\Delta x/2l$.
The same result can be obtained from the continuum limit of this random walk. Although the regime $k s(x)\ll 1$ does not hold over the entire space, we will show that this is indeed the exact solution of the process at long times, after the stationary equilibrium state has been reached. 
\subsection*{Continuum Limit and the Itô Interpretation}

To move from the discrete recurrence relation to a continuous description, we establish that the process is governed by a stochastic differential equation (SDE). We argue that the Itô interpretation is the uniquely appropriate framework for this model because the discrete update rule, $n_{t+1} = n_t + \sigma_t s(n_t)$, evaluates the position-dependent step size $s(n_t)$ strictly at the \textit{prepoint} position $n_t$ before the stochastic increment $\sigma_t$ is applied. This causal structure is the defining feature of Itô calculus and ensures the process remains fundamentally Markovian, in which the future state depends only on the current configuration.

In the continuum limit, the evolution of the walker's position $x_t$ can be described by:
\begin{equation}
     dx_t = C\left(\frac{|x_t|}{l\tau} + rV_0\right) dW_t\,,
    \label{eqcont}
\end{equation}
where $C$ is a constant that ensures the proper dimensions, $W_t$ is a Wiener process, $V_0\equiv \Delta x/\tau$, and $l$ and $r$ were defined previously. Because this SDE has dimensions of length and $dW_t$ scales as $\sqrt{dt}$, where $dt$ is an infinitesimal time increment, the proper scale must be included; thus, $C\left(\frac{|x_t|}{l\tau} + rV_0\right) dW_t\sim L$, where $L$ denotes the unit of length. Moreover, $\langle dW_t dW_{t'}\rangle=dt$ for $t=t'$ and $0$ otherwise, and $\langle dW(t)\rangle=0$. Therefore, we consistently choose

\begin{equation}
     dx_t = \sqrt{2l\tau}\left(\frac{|x_t|}{l\tau} + rV_0\right) dW_t\,.
    \label{eqcont}
\end{equation}
\begin{figure}%[htb]
    \centering
    \includegraphics[width=12cm]{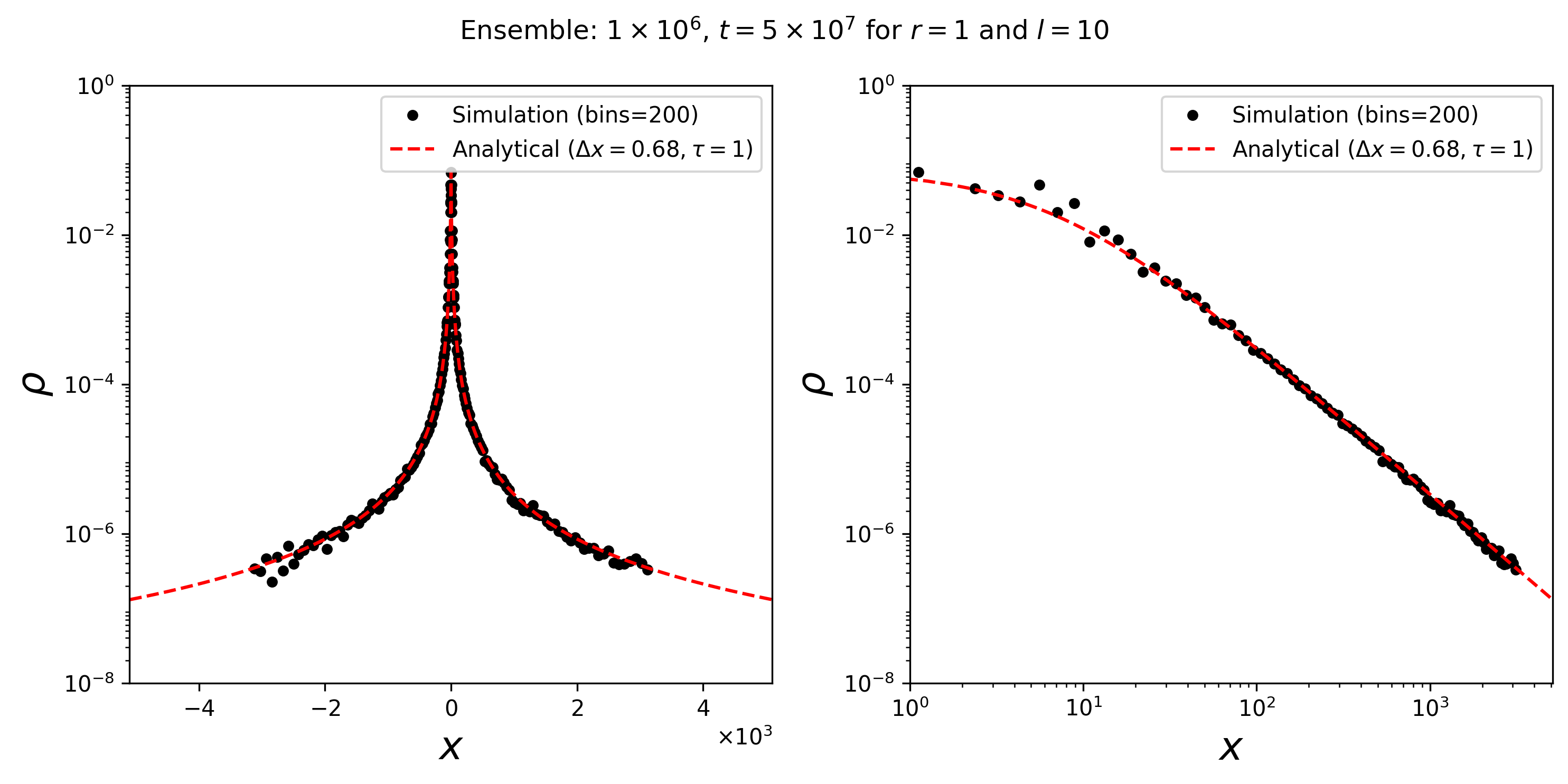} 
   
    \includegraphics[width=12cm]{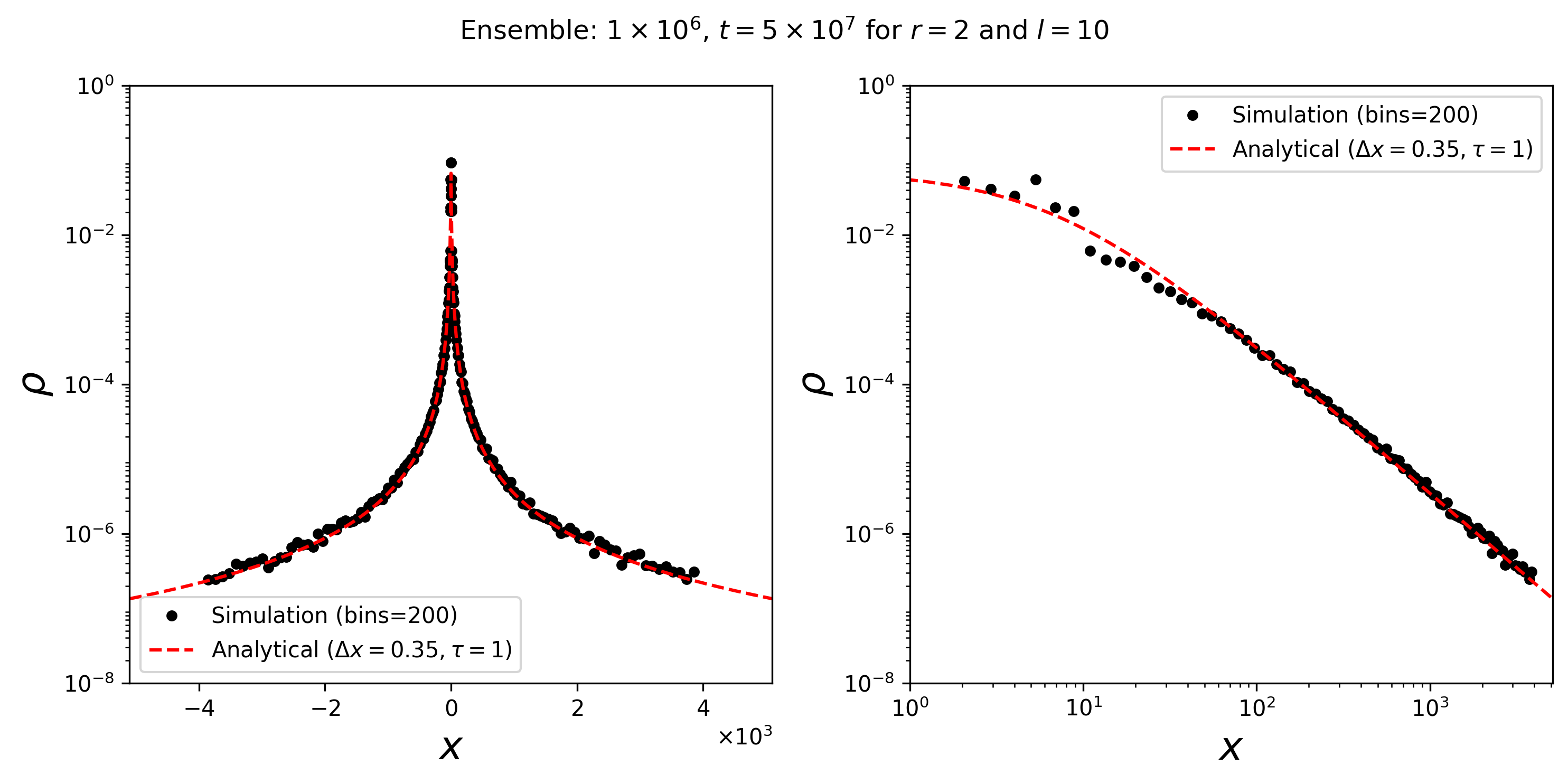}
    \includegraphics[width=12cm]{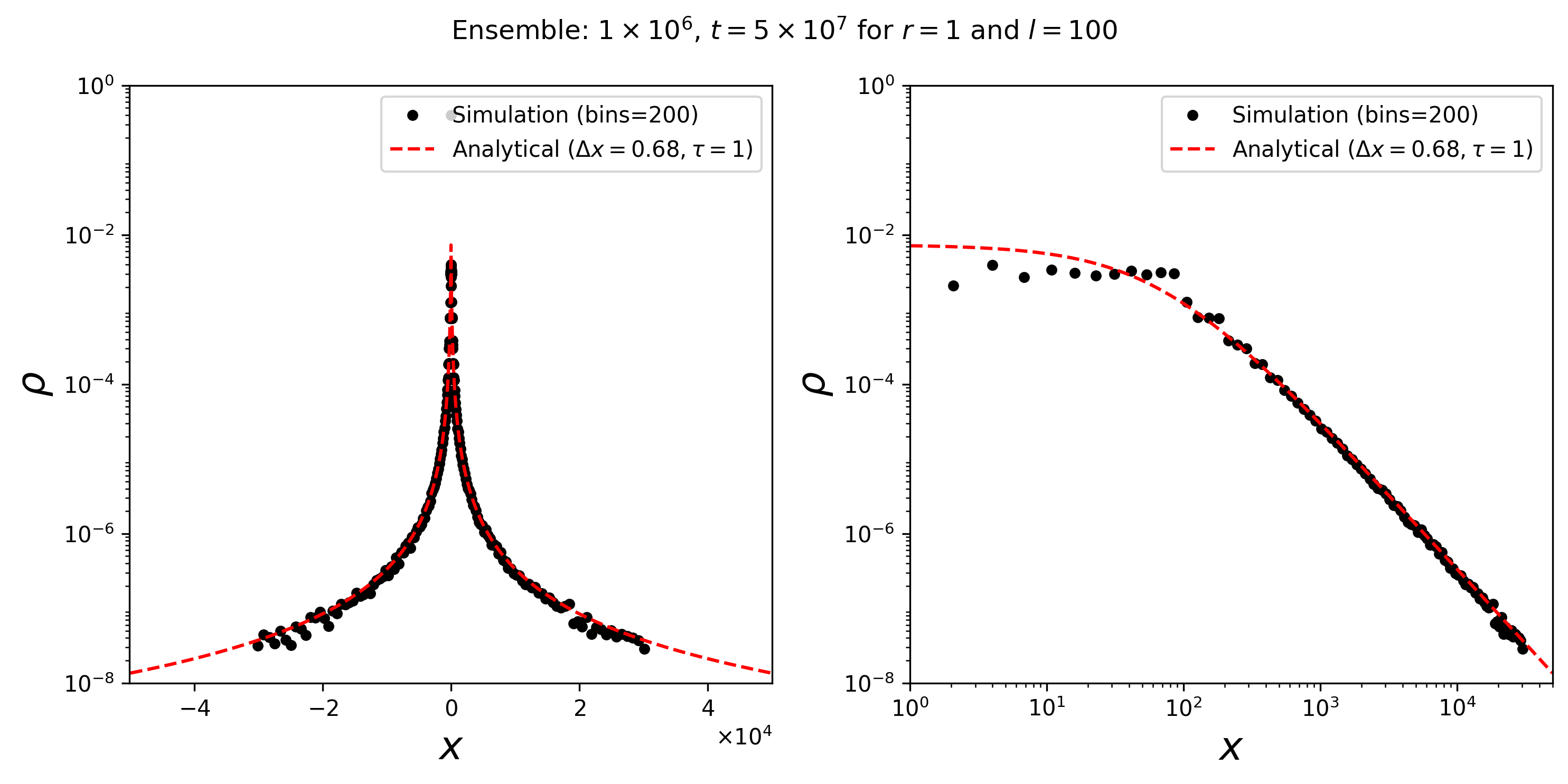}   
    \caption{\textit{Top left and right}: $\rho$ versus $x$ for $r=1$ and $l=10$ on log-linear and log-log scales, respectively. \textit{Center left and right}: $\rho$ versus $x$ for $r=2$ and $l=10$ on log-linear and log-log scales, respectively. \textit{Bottom left and right}: $\rho$ versus $x$ for $r=1$ and $l=100$ on log-linear and log-log scales, respectively.}
    \label{fig1}
\end{figure}

In the Itô sense, we define a transformation $y_t \equiv y(x_t)$ following Itô's Lemma: $dy_t = y_t'(dx_t) + \frac{1}{2}y_t''(dx_t)^2$. By choosing the transformation such that $y_t' = 1/\left(\frac{|x|}{l\tau} + rV_0\right)$, we map the multiplicative noise into an additive noise regime:
\begin{equation}
    dy_t = 
    \begin{cases}
    -dt + \sqrt{2l\tau}\,dW_t & x_t > 0 \\
    dt +  \sqrt{2l\tau}\,dW_t & x_t < 0.
    \end{cases}
\end{equation}
Let us emphasize that the arbitrary integration constant in $y$ is chosen appropriately so that the argument of the logarithm becomes dimensionless. This leads to two Fokker-Planck equations for the probability density $\tilde{\rho}$ in $y$-space:
\begin{equation}
    \partial_t \tilde{\rho} = -\partial_y J_{\pm}, \quad J_{\pm} \equiv \mp \tilde{\rho} - l\tau\partial_y \tilde{\rho}.
\end{equation}
In the equilibrium stationary state, setting the flux to zero ($J_{\pm} = 0$) and the transformation back to the original coordinates via $\rho_{st}(x) = \tilde{\rho}_{st}(y(x)) |y'|$ yields
\begin{equation}
    \rho_{st}(x) = 
    \begin{cases}
    \tilde{\rho}_0 \frac{1}{\left(\frac{x}{l\tau} + rV_0\right)^2} & x_t > 0 \\
    \tilde{\rho}_0 \frac{1}{\left(-\frac{x}{l\tau} + rV_0\right)^2} & x_t < 0\,,
    \end{cases}
\end{equation}
which simplifies to the following normalizable, symmetric distribution:
\begin{equation}
    \rho_{st}(x) = \frac{\rho_0}{\left(\frac{|x|}{l\tau} + rV_0\right)^2}\,,
    \label{eqdist}
\end{equation}
where $\rho_0 = rV_0 / (2l\tau)$ is the normalization constant. This analytical solution, characterized by power-law tails decaying as $|x|^{-2}$, matches the long-time behavior observed in our extensive numerical simulations for some values of $l$ and $r$ (see Fig.\ref{fig1}). In particular, for $l=100$, the analytical result agrees with the numerical simulations over six decades in $x$. If we write the solution in terms of $\Delta x$ and $\tau$, we recover the previous expression derived from the generating function. Notice that $V_0$ is indeed a type of velocity independent of the value of $r$; however, it is useful to define the  velocity $V\equiv rV_0$, which is proportional to $r$. The acceleration enters the normalization constant, since $|a| \equiv V/(l\tau)=rV_0/(l\tau)$. In particular, for $l\to \infty$, we recover a constant, non-normalizable distribution. Thus, we recover a simple random walk, which has no stationary distribution or acceleration. We also recover a trivial, non-normalizable equilibrium distribution if $r=0$, which means that no initial acceleration exists ($|a|=rV_0/(l\tau)=0$); this case is related to a purely transient log-normal process \cite{Shneidman1998}. Although velocity and acceleration lack their traditional kinematic interpretation in a purely symmetric random walk, they acquire physical significance along the shortest fluctuation-free paths. Rather than being unphysical artifacts, the effective roles of these quantities can be rigorously derived and interpreted within the framework of the Onsager-Machlup formalism~\cite{OnsagerMachlup1953I, OnsagerMachlup1953II,Paraguassuetal2021}.

\subsection*{The Onsager-Machlup approach} 
Let us consider the Onsager-Machlup functional for our continuous process described above. It can be written as
\begin{equation}
    \mathcal{P}(y,t|y_0,0)=\int\limits_{y(0)=0}^{y(t)=y} D\,y \,e^{-S[y]}\,,
\end{equation}
where $S[y]=\frac{1}{4l\tau}\int_{0}^t dt' \left(\dot{y}(t')+\sgn(y(t'))\right)^2$. The trajectory that minimizes this action is associated with the effective Lagrangian $L(y,\dot{y},t)=\frac{1}{4 l\tau}\left(\dot{y}(t)+\sgn(y(t))\right)^2$ for the \textit{classical trajectories}, namely $y_{cl}$. Therefore, we have
\begin{equation}
    \frac{d}{dt}\left(\frac{\partial}{\partial \dot{y}_{cl}}L\right)=\ddot{y}_{cl}/(2l\tau)=0\,.
\end{equation}
Thus, under the boundary conditions $x(0)=0$, which implies
\begin{equation*}
    |y_{\text{cl}}(0)| = l\tau \ln\left[ \frac{\frac{|x_{\text{cl}}(0)|}{l\tau} + V}{V} \right] = 0\,,
\end{equation*} 
and $\dot{y}_{\text{cl}}(0) = -\operatorname{sgn}(y)$ for trajectories moving from an 
arbitrary region back to the origin, we have

\begin{equation}
    y_{cl}(t)=y_{cl}(0)-\sgn(y)\, t\,.
\end{equation}
Then, in the $x$ variable, this yields

\begin{equation}
    x_{cl}(t)=\sgn(x)\, l\tau V\left[e^{-t/(l\tau)}-1\right]\,,
\end{equation}
where $V\equiv rV_0$.
From the results above, we finally obtain, in the limit of $t\to\infty$,
\begin{equation}
    |x_{cl,\infty}|=l\tau V,
\end{equation}
which means that $V=|x_{cl,\infty}|/(l\tau)$. The acceleration is then obtained as $|a|=|x_{cl,\infty}|/(l^2\tau^2)$. Therefore, the velocity $V$ and acceleration $a$ are related to the terminal position of the action-minimizing trajectory, which tends to return to the origin; they are also related to the threshold $l$ and relaxation time $\tau$. Notice that $|x_{cl,\infty}|$ yields exactly $l\,\tau\, r\,V_0=l\,r\Delta x$, which, for $r=1$, corresponds to the region where the crossover between simple and accelerated walks occurs. However, for $r>1$, $|x_{cl,\infty}|$ increases as $r$ increases. The probability distribution $\rho_{st}(0)$ at the origin is $\rho_{st}(0)=1/(2 |x_{cl,\infty}|)\propto 1/|x_{cl,\infty}|$. This implies that $V$ and $a$ represent, respectively, the effective velocity and acceleration that initially drive the walker away from the origin. 

We now consider the classical trajectories moving far away from the origin. In this case, we simply change $t\to-t$, which yields

\begin{equation}
    |x_{cl}(t)|= l\tau V\left[e^{t/(l\tau)}-1\right]\,.
\end{equation}
This implies that $\left[\left(\frac{|x_{cl}|}{l\tau}+V\right)/V\right]=e^{t/(l\tau)}$; therefore, we conclude that $v(x)=|x|/(l\tau)+V$ and $a(x)= |x|/(l^2\tau^2)+V/(l\tau)$, so that, when $|x|/(l\tau)\ll V$, we have $v(x)\approx V$ and $a(x)\approx V/(l\tau)$. For this particular classical trajectory, we have $S[y]=t/(l\tau)=|y|/(l\tau)$; therefore $\mathcal{P}(y)= e^{-|y|/(l\tau)}$, which recovers, after the change of variable $y\to x$, the equilibrium distribution in Eq. \eqref{eqdist}.

A similar result can be obtained from the discrete form of the process. Let us consider an arbitrary variation of the position, $\Delta x_t$, as

\begin{equation}
    \Delta x_t=\tau \sigma_t \left(|x_t|/(l\tau)+V \right).
\end{equation}
Thus, we define the velocity $v_t$ associated with this motion as $v_t\equiv\Delta x_t/\tau=\sigma_t \left(|x_t|/(l\tau)+V \right)$.
The corresponding velocity increment, $\Delta v_t$, is
\begin{equation}
    \Delta v_t =\sigma_{t+1}\left(\frac{|x_{t+1}|}{l\tau}+V\right)-\sigma_t \left(\frac{|x_{t}|}{l\tau}+V \right)\,.
\end{equation}
Since $x_{t+1}=x_{t}+\tau \sigma_t \left(\frac{|x_{t}|}{l\tau}+V\right)$, we have

\begin{equation}
   \Delta v_t =\sigma_{t+1}\left(\frac{\left|x_{t}+\tau \sigma_t \left(\frac{|x_{t}|}{l\tau}+V\right)\right|}{l\tau}+V\right)-\sigma_t \left(\frac{|x_{t}|}{l\tau}+V \right) .
\end{equation}
Since we are interested in the behavior of the shortest fluctuation trajectories, we restrict our analysis to the cases where $\sigma_{t+1}=\sigma_{t}=+ 1$. Therefore, we obtain

\begin{equation}
    \Delta v_t =\left(\frac{x_t}{l^2 \tau}+\frac{V}{l}\right)\,,
\end{equation}
where for $x_t/(l \tau)\ll V$, the expression becomes $\Delta v_t/\tau\sim V/(l\tau)=+a$. The result for $x_t<0$ can be derived similarly, so the general expression for long walks is determined by $\Delta v_t/\tau \sim |x_{cl,\infty}|/(l^2 \tau^2)=|a|$. Thus, we conclude that velocity and acceleration in this process are not unphysical artifacts, but rather inherent characteristics of the walkers' shortest trajectories. In this context, for the simple random walk ($l\to \infty$), we always have $\Delta x_t/\tau=V$ and $|a|=0$, while for a log-normal process, we have $\Delta x_t/\tau~\sim v_t\propto x_t$ and $a_t\propto x_t$, with $|a|=0$ at all time intervals. Neither case  possesses heavy tails, and both have $|a|=0$. Therefore, the initial acceleration $|a|\neq 0$ that pushes the walker away from the origin is the underlying mechanism behind both the emergence of asymptotic power-law distributions at equilibrium and the existence of normalizable stationary distributions.

\section*{Final remarks}

In summary, we have introduced and analyzed a one-dimensional, discrete-time accelerated random walk that exhibits a novel mechanism for the emergence of heavy-tailed statistics. Through analytical derivations in the continuum limit and extensive numerical simulations, we 
demonstrated that a simple, deterministic, position-dependent step rule generates a robust, strongly non-Gaussian stationary distribution. The agreement between simulations and the analytical solution, $\rho_{\text{st}}(x) \propto (|x|/l+r\Delta x)^{-2}$, confirms asymptotic power-law tails decaying as $\rho_{\text{st}} \sim |x|^{-2}$.

A key finding is that these scale-free statistics emerge from a deterministic, local feedback rule rather than from \textit{a priori} stochastic assumptions. Unlike L\'evy flights, which require prescribed heavy-tailed jump distributions, or continuous-time random walks, which often rely on 
non-Markovian memory, our model is strictly Markovian. The heavy tails are an emergent property of the dynamics, driven by a positive feedback loop where the step size $s(n_t)$ grows with distance from the origin.

Remarkably, power-law statistics with the same exponent obtained in our results are repeatedly observed in a wide range of complex systems, including wealth and income distributions~\cite{dragulescu2001}, solar and stellar flares~\cite{aschwanden2022}, extinction event
sizes~\cite{newman1996}, material-failure avalanches~\cite{pradhan2006}, ecological size spectra~\cite{camacho2001}, linguistic frequency
patterns~\cite{ferrer2001}, and scale-free or self-similar
networks~\cite{barabasi1999,adamic2000,serrano2012}. This recurrence across social, biological, and physical domains underscores the importance of identifying minimal, analytically solvable mechanisms that can produce specific asymptotic exponents without invoking long-range memory or intrinsic
disorder. Our Markovian, position-dependent walk provides such a minimal mechanistic pathway, offering a new microscopic basis for the emergence of the $-2$ power law in the asymptotic limit of the probability distribution.

The derived stationary distribution implies an undefined variance, consistent with the non-Gaussian nature of the process. This analytically tractable framework provides new insights into power-law emergence in systems where 
mobility or activity is environmentally dependent, such as diffusion in heterogeneous media or financial models with wealth-dependent volatility. Future work will explore transient dynamics, moment scaling, generalizations 
of the acceleration rule (e.g., $s(x) \sim |x|^\beta$), and extensions to higher dimensions, which may reveal new universality classes of anomalous transport.

\section*{Acknowledgments}
We acknowledge fruitful discussions with M. A. Pires, M. M. R. P. Martins, and  M. Calvelli as well as partial financial support from
Conselho Nacional de Desenvolvimento Científico e Tecnológico
 (CNPq), Fundação Carlos Chagas Filho de Amparo à Pesquisa do Estado do Rio de Janeiro (FAPERJ), and Coordenação de Aperfeiçoamento de Pessoal de Nível Superior (CAPES) (Brazilian agencies). 

%\appendix

%\section*{Appendixes}

\end{document}